\documentclass[varenna,cite]{cimento}

\usepackage{epsfig}

\begin{document}

\newcommand{\oneStwoS}{\mbox{$1S$-$2S$}}
\newcommand{\oneSoneS}{\mbox{$1S$-$1S$}}
\newcommand{\Hdown}{\mbox{$H\!\!\downarrow$}}
\newcommand{\Hup}{\mbox{$H\!\!\uparrow$}}
\newcommand{\Lalpha}{\mbox{Lyman-$\alpha$}}

\varinfo{140}{1998}{}

\title{Bose-Einstein Condensation of Atomic Hydrogen}

\author{Daniel~Kleppner, Thomas~J.~Greytak, Thomas~C.~Killian, 
Dale~G.~Fried, Lorenz~Willmann, David~Landhuis, \atque Stephen~C.~Moss}

\shortauthor{Daniel~Kleppner, Thomas~J.~Greytak, \etal}

\institute{Department of Physics
Massachusetts Institute of Technology
Cambridge, Massachusetts 02139}

\maketitle

\section{INTRODUCTION}

Bose-Einstein condensation in atomic hydrogen was observed for the
first time just a few weeks before this session of the Enrico Fermi
School, and so these lectures constitute a somewhat breathless first
report.  However, the search for Bose-Einstein condensation (BEC) in
hydrogen began many years ago, and it has a long and colorful history.
Out of that history emerged an extensive body of knowledge on the
behavior of hydrogen at low temperatures that provided the foundation
for achieving BEC in hydrogen.  Much of it has been described in
reviews \cite{gkl84,swa84,bhk86,gre95}.  Consequently, we will dwell
on only those features of that research that provide essential
background, and concentrate on the most recent developments in which
BEC in hydrogen advanced from a long period of being tantalizingly
close, to being real.

\section{ORIGINS OF THE SEARCH FOR BEC IN AN ATOMIC GAS}

In April, 1976, W. C. Stwalley and L. H. Nosanow published a letter
summarizing studies on the equation of state of spin-polarized
hydrogen, \Hdown\ \cite{sno76}.  Because there are no bound states of
molecular hydrogen in the triplet state, \Hdown\ behaves like a simple
monatomic gas, but a gas with a remarkable property.  Because of the
weak \Hdown~-~\Hdown\ potential and the atom's low mass, \Hdown\
remains a gas at temperatures down to $T = 0$.  Consequently, it might
be possible to cool \Hdown\ to the quantum regime and achieve BEC.
That paper essentially launched the search for BEC in an atomic gas.
It triggered a flurry of experiments and enough activity for an entire
session of the December, 1978, APS meeting to be devoted to the
stabilization of hydrogen \cite{APS}.

The critical density for the BEC transition in a non-interacting gas
is $n_{\rm c} =2.612 \Lambda_{\rm T}^{-3}$, where $\Lambda_{\rm T} =
\sqrt{ 2 \pi \hbar^2/k_{\rm B} T m}$ is the thermal de Broglie wavelength.
Because $n_{\rm c}$ depends on the product of temperature and mass, for a
given density hydrogen condenses at a higher temperature than any
other atom.  Hydrogen offered two other attractions as a
candidate for BEC.  The hydrogen atom is generally appealing for basic
studies because its structure and interactions can be calculated from
first principles.  Furthermore, \Hdown\ constitutes a nearly ideal Bose
gas: it has an anomalously small $s$-wave scattering length and its
interactions are weak.

In the years since the search for BEC in atoms started, there was
a revolution in techniques for cooling and manipulating atoms using
laser-based methods that culminated in the achievement of BEC in alkali
metal atoms \cite{JILA,dma95,bst95}.  In light of these advances,
hydrogen's special attractions must be viewed from a new
perspective.  Spin-polarized hydrogen's unique property of remaining a
gas at $T = 0$ is evidently not essential for BEC.  Although all
species except helium should be solid at sub-kelvin temperatures,
laser-cooled atoms do not even liquefy because they never hit
surfaces.  In the absence of surface collisions, a gas to liquid
transition requires three-body collisions to initiate nucleation.  At the
densities used to achieve BEC in alkali metal atom systems, however,
the three-body recombination rate is so low that it can be neglected
at all but the highest densities.  Hydrogen's relatively high
condensation temperature is also not a crucial advantage.  Laser
cooling methods make it is possible to cool alkali metal gases to the
microkelvin regime, far colder than possible by conventional cryogenic
means.  Once the atoms have achieved the laser-cooling temperature
limit, they can be efficiently cooled into the nanokelvin regime by
evaporative cooling.  Finally, hydrogen's close to ideal behavior must
now be regarded as a serious experimental disadvantage.  This is
because all routes to BEC used so far involve evaporative cooling.
Evaporation requires collisions for maintaining thermal equilibrium as
the system cools.  Because of hydrogen's small scattering length, its
collision cross section is tiny and evaporation is much slower than in
other systems.

Nevertheless, now that hydrogen can be Bose-Einstein condensed, its
simplicity continues to give the atom unique interest.  The
techniques, condensate size, and the general conditions for
condensation are different from those of other realizations and one
can expect that this new condensate will open complementary lines of
research.

Spin polarized hydrogen is created by the magnetic state selection of
hydrogen at cryogenic temperatures.  In high magnetic fields the
electron and proton spin quantum numbers are $m_e = -\frac{1}{2},~m_p
= \pm \frac{1}{2}$ for \Hdown\, and $m_e = +\frac{1}{2},~m_p = \pm
\frac{1}{2}$ for \Hup.  The governing parameter in magnetic state
selection is the interaction energy in a strong magnetic field, $B$.
In temperature units, this is $T_o = \mu_{\rm B} B / k_{\rm B} = 0.67(
B/ {\rm tesla})$~K.  Here $\mu_{\rm B}$ is the Bohr magneton, and
$k_{\rm B}$ is Boltzmann's constant.  For a field of $10$~T, which is
readily achieved in a superconducting solenoid, $T_o = 6.7$~K. At a
temperature of $0.3$~K the ratio of densities $n$(\Hup) / n(\Hdown)
$\sim \exp(- 2T_o /T)$, which is about $10^{-20}$.  The spin
polarization is essentially 100\%.

Another crucial experimental parameter in creating \Hdown\ is the
binding energy for adsorption on a liquid helium surface,
$E_{\rm b}$.  Hydrogen must make collisions with a cold surface in
order to thermalize at cryogenic temperatures, but if the atoms become
adsorbed the gas will rapidly recombine.  In thermal
equilibrium, the surface density $\sigma$ and volume density $n$ of
\Hdown\ are related by $\sigma = \Lambda_{\rm T} n \exp{(-E_{\rm
b}/k_{\rm B} T)}$.  For hydrogen-helium, $E_{\rm b} = 1$~K.
Consequently, for temperatures below about 0.1~K the atoms move to the
walls where they can recombine by two- and three-body processes.  The
initial searches for BEC were in the temperature regime $0.2 \sim
0.7$~K.  The critical density at a temperature of $0.5$~K is $n_{\rm
c} = 6.8\times 10^{19}~{\rm cm}^{-3}$.  

Spin-polarized hydrogen was first stabilized by I. F. Silvera and J.
T. M. Walraven in 1980 \cite{swa80}, and magnetically confined by the
M.I.T. group \cite{cgk81}.  In these experiments the field of a
superconducting solenoid provided both state-selection and axial
confinement.  Radial confinement was provided by a superfluid
helium-coated surface.

The highest density achieved with \Hdown\ under controlled conditions
was $4.5 \times 10^{18}~{\rm cm}^{-3}$, at a temperature of $0.55$~K
\cite{bhk86}.  The density was limited by three-body recombination, a
process that had been predicted by Kagan \etal\ \cite{kvs80}.  Because
the heat generation in three-body recombination increases as the cube
of the density, at higher density the gas would essentially
self-destruct.  An alternative route to BEC was required.

\section{HYDROGEN TRAPPING AND COOLING}

The new route toward BEC led to temperatures in the microkelvin regime
where $n_{\rm c}$ is so low that three-body recombination is
unimportant.  At temperatures much below $0.1$~K, however, surface
adsorption and recombination become prohibitive.  To avoid surfaces,
Hess suggested strategies for confining \Hup\ atoms (in the ``low
field seeking'' states) in a magnetic trap, and cooling the gas by
evaporation \cite{hes86}.

Trapping and cooling requires an irreversible process for losing
energy.  In laser-cooling and trapping experiments, this process is
spontaneous emission.  Unfortunately, laser methods are not well
suited to hydrogen.  Aside from the lack of convenient light sources,
the laser cooling temperature limit for hydrogen is relatively high.
The limit is determined by the recoil energy for single photon
emission, and in hydrogen it is more than a millikelvin.  The method
proposed by Hess employed elastic scattering for both trapping and
cooling.

The magnetic trap proposed by Hess consists of a long quadrupole field
to confine the atoms radially with axial solenoids at each end to
provide axial confinement---an elongated variant of the
``Ioffe-Pritchard'' configuration \cite{pri83}. The Ioffe-Pritchard
potential is
\begin{equation}
V({\bf r})=\sqrt{(\alpha \rho)^2 + (\beta z^2 +\theta)^2}-\theta ,
\end{equation}
with radial potential gradient $\alpha$, axial curvature $2\beta$, and
bias energy $\theta$.  At low energies the potential is quasi-harmonic
with radial and axial oscillation frequencies
\begin{eqnletter}
\label{osc.freq.eqn}
\omega_\rho &=& \frac{\alpha}{\sqrt{m (\beta z^2+\theta)}} \\
\omega_z &=& \sqrt{\frac{2\beta}{m}}
\end{eqnletter}
The trapping fields are produced inside a cell that confines the gas
cloud while atoms are loaded into the trap.  The walls of the
helium-coated cell are held at a temperature of about 275~mK.  The
cell is filled with a puff of hydrogen and helium from a low
temperature discharge and the walls of the cell are then cooled,
making them ``sticky.''  Atoms with high energy leave the trap and
stick to the walls.  Thermal contact between the walls and the trapped
gas is broken because these atoms are unable to return.  The trapped
gas cools by evaporation and reaches an equilibrium temperature of
about 1/12th the trap depth.

To further cool the gas, Hess proposed a process of forced
evaporation.  The height of the potential barrier is slowly decreased,
allowing energetic atoms to escape and thus reducing the average
energy of the system.  As the system's reduced energy is redistributed
by collisions, the temperature falls.  Unlike ordinary evaporation
which has a temperature limit determined by the vapor pressure of the
material, forced evaporation can be continued to almost arbitrarily
low temperatures.  The process is surprisingly efficient because the
escaping atoms carry away a great deal of energy, $(5 \sim 10) k_{\rm
B} T$ per atom.

Spin-polarized hydrogen was confined by a pure magnetic trap in 1987
by the M.I.T. group \cite{hkd87} and also in Amsterdam \cite{wal}.
Shortly thereafter, in the first demonstration of forced evaporative
cooling the M.I.T. group achieved a temperature of 3~mK \cite{mds88}.

\begin{figure}[t]
\centering\epsfig{file=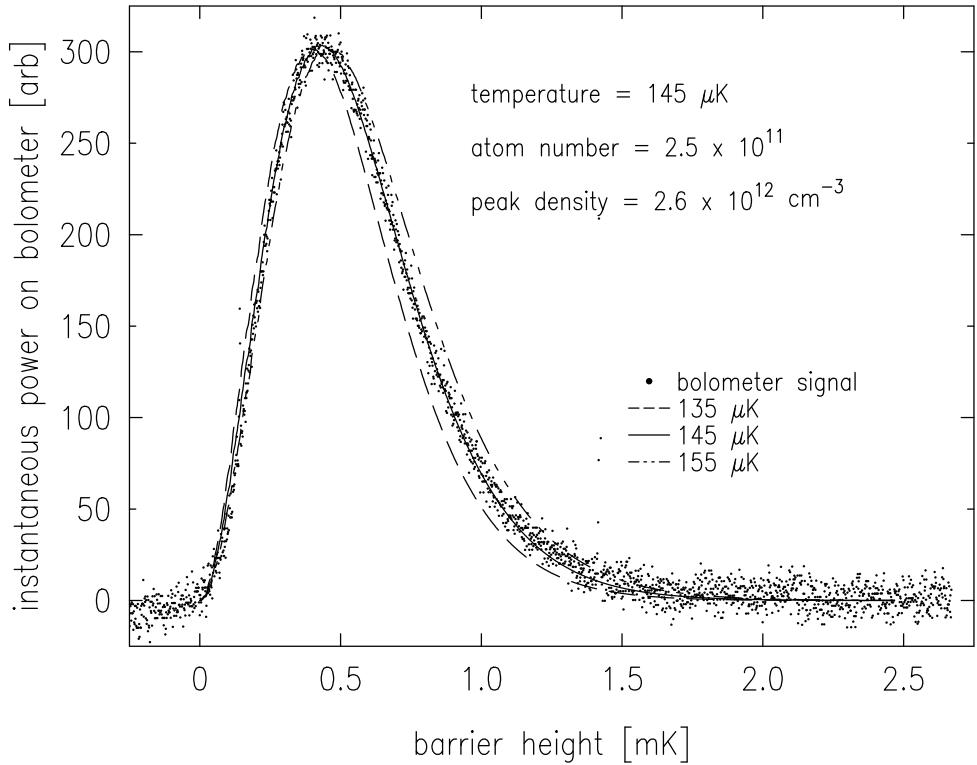}
\caption{ Energy distribution of trapped \Hup\ after evaporative
cooling: plot of the atom flux as the trap barrier is rapidly lowered.
The atom flux is measured by detecting the molecular recombination
heat on a sensitive bolometer.  Calculated distributions for three
temperatures are shown.}
\label{trap.dump.fig}
\end{figure}

In all of these experiments, the atoms were studied by monitoring the
hydrogen flux as the atoms were dumped from the trap.  With this
technique \cite{mds88,dsm89}, the field of one of the axial confining
coils is reduced, allowing the atoms to escape from the trap.  Once
out of the trap, the hydrogen rapidly adsorbs on the walls of the cell
and recombines.  This heat of molecular recombination---4.6~eV per
event---is measured by a small bolometer within the cell but outside
the trap.  If the confinement field is reduced rapidly compared to the
thermalization time, then the atom flux as a function of barrier
height reveals the energy distribution of the gas.  From this the
temperature can be inferred.  A typical energy distribution is shown
in fig.\ \ref{trap.dump.fig}.
Integrating the signal gives a measure
of the total number $N$ of atoms trapped.

\begin{figure}[t]
\centering\epsfig{file=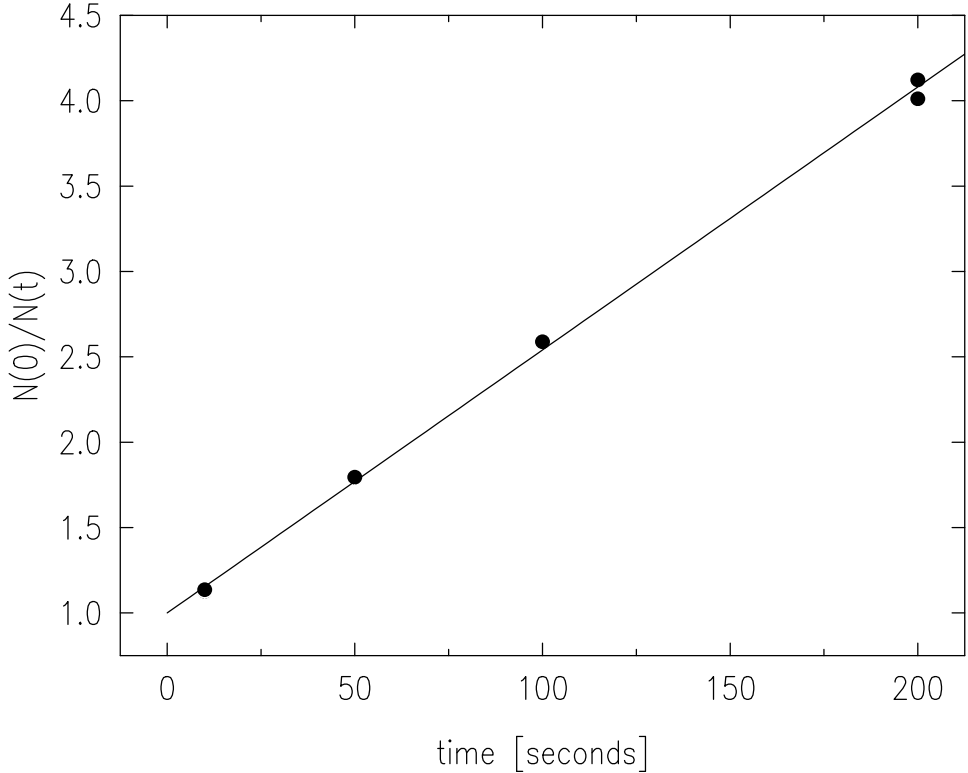}
\caption{ Determination of the sample density by observing decay due
to dipolar relaxation.  Five samples were indentically prepared and
then held for varying times (horizontal axis) before being dumped from
the trap.  The integrated recombination signal from the dump gives the
number of atoms in the trap, and is plotted on the vertical axis.  The
straight line fit indicates a density of $6.0\times 10^{13}~{\rm
cm}^{-3}$.  From \cite{kfw98} }
\label{density.measurement.fig}
\end{figure}

The density can also be determined by observing the decay of the
trapped gas.  The primary decay mechanism in \Hup\ is dipolar
relaxation, a process in which the spin angular momentum of a pair of
atoms is transferred to their orbital angular momentum, while one or
both of the atoms makes a transition to an untrapped state and
escapes.  Because dipolar relaxation is a two-body process, the
density decays according to $\dot{n} = - g n^2$, where the dipolar
decay constant $g$ has been calculated \cite{lsv86,skv88} and also
measured \cite{hkd87,wal}.  The calculated value is $g=1.2\times
10^{-15}~{\rm cm}^3 \rm s^{-1}$ \cite{gterms.ref}.  The total sample
decays according to $\dot{N}= -\kappa g N$ where
\begin{equation}
\kappa = \frac{\int \exp\left[-2V({\bf r})/k_{\rm B} T\right]dV}
{\int \exp\left[-V({\bf r})/k_{\rm B} T\right]dV}.
\end{equation}
For dipolar decay the number of trapped atoms decays according to 
\begin{equation}
N(0)/N(t)=1+\kappa g n_o t .
\end{equation}
Hence, the initial peak density $n_o$ can be extracted from a plot of
$N(0)/N(t)$.  An example of such a plot is shown in fig.\
\ref{density.measurement.fig}.  
Typically $\kappa\sim 1/5$ and the
characteristic decay time at a density of $10^{14}~{\rm cm}^{-3}$ is
40~s.

Dipolar relaxation takes place predominantly where the density is
high.  This is at the minimum of the trap where the mean energy is
low.  Just as evaporation cools by removing the most energetic atoms,
dipolar relaxation heats by removing the least energetic atoms.
Consequently, the trapped gas comes to a thermal equilibrium in which
cooling due to evaporation is balanced by heating due to dipolar
relaxation.

These attempts to reach BEC in \Hup\ stopped short of the quantum
degenerate regime by about a factor of six in phase space density
\cite{dsy91}.  Several complications arose.  The diagnostic technique
of dumping the atoms out of the trap began to give ambiguous results
because the energy distributions of the samples would change
significantly during the dumping process.  The sensitivity of the
detection process was not sufficient.  The efficiency of the
evaporative cooling process degraded significantly, precluding further
cooling.  Finally, even if the quantum degnerate regime could be
reached, a condensate would degenerate during the sample dumping
process, and would have been unobservable.  These complications
required a new diagnostic technique and a more efficient evaporation
process.

\section{OPTICAL DETECTION OF TRAPPED HYDROGEN}

Trapped hydrogen can be detected {\em in situ} by photoabsorption
using one- and two-photon transitions.  The Amsterdam group has used
both methods.  Using a \Lalpha\ light source they observed absorption
of the principal transition \cite{swl93}.  However, the absorption
spectrum has a large natural linewidth and displays structure due to
the inhomogeneous magnetic field, limiting its use for analyzing
momentum.  They also observed the $1S$-$3S$ and $1S$-$3D$ transitions
using resonantly enhanced two-photon excitation exploiting a virtual
state near $2P$ \cite{pmw97}.  The $1S$-$3S$ transition is insensitive
to magnetic fields and is potentially well suited for analyzing
momemtum at temperatures down to 1~$\mu$K.  The $1S$-$3D$ transition
has a matrix element that is ten times larger, but a natural linewidth
that is ten times broader.

We have employed two-photon Doppler-free spectroscopy of the
\oneStwoS\ transition, using two-photons from a single laser tuned to
243~nm, twice the \Lalpha\ wavelength.  The spectrum is essentially
unperturbed by the magnetic field.  This method provides excellent
momentum resolution but suffers from the low excitation rate of a
forbidden transition.  In two-photon spectroscopy using a single light
source normally only the narrow Doppler-free spectrum is observed,
excited by counter-propagating laser beams that eliminate the first
order Doppler effect.  However, at very low temperatures in our
experiment the broad Doppler-sensitive spectrum is also visible,
excited by absorbing two photons from the same laser beam.  Because
the momentum distribution in a condensate is much narrower than in a
normal gas at the same temperature, the Doppler-sensitive spectrum is
well suited to probing the condensate.  In addition to its application
to observing BEC, Doppler-free spectroscopy of cold trapped hydrogen
has a potential application to the precision spectroscopy of hydrogen,
in which the \oneStwoS\ transition plays a central role \cite{uhg97}.
Under suitable conditions, time-of-flight broadening disappears and a
spectral resolution close to the natural linewidth of 1.3~Hz should be
possible \cite{kle89}.

\begin{figure}[t]
\centering\epsfig{file=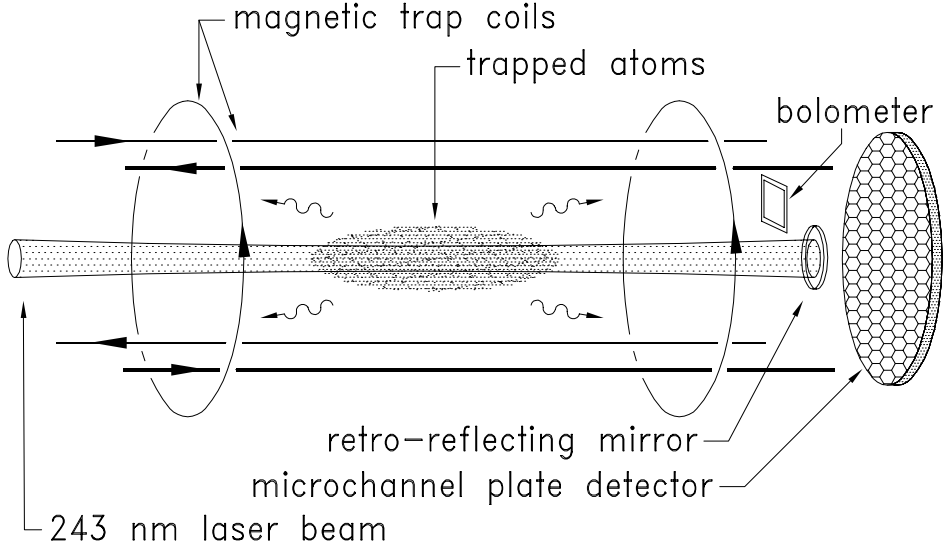}
\caption{ Schematic diagram of the apparatus.  The superconducting
magnetic coils create trapping potential that confines atoms near the
focus of the 243~nm laser beam.  The beam is focused to a 50~$\mu$m
waist radius and retroreflected to produce the standing wave required
for Doppler-free two-photon absorption.  After excitation,
fluorescence is induced by an applied electric field.  A small
fraction of the 122~nm fluorescence photons are counted on a
microchannel plate detector.  The bolometer is used for
diagnostic purposes.  Not shown is the trapping cell which surrounds
the sample and is thermally anchored to a dilution refrigerator.  The
actual trap is longer and narrower than indicated in the diagram.}
\label{apparatus.spectroscopy.fig}
\end{figure}

In our method \cite{cfk96}, the atoms are excited by a pulse
of 243~nm radiation.  For observing BEC the pulse length is typically
0.4 ms, but for spectroscopic studies pulses as long as 5~ms have been
used.  (In the absence of an electric field we have observed that the
$2S$ atoms live for close to the natural lifetime, 122~ms
\cite{cfk96}.)  Because of the low excitation rate, it is not feasible
to observe photoabsorption.  Instead, the excited atoms are detected
by switching on an electric field which Stark-quenches the $2S$ state,
mixing it with the $2P$ state which promptly decays.  The emitted
\Lalpha\ photon is detected.  A diagram of the apparatus is shown in
fig.\ \ref{apparatus.spectroscopy.fig}.
The energy equation for two-photon excitation of an atom from state
$i$ to $f$, with initial momentum ${\bf p}_i$ and final momentum ${\bf
p}_f = {\bf p}_i + \hbar({\bf k}_1 +{\bf k}_2)$, where ${\bf k}_1$ and
${\bf k}_2$ are the wave vectors of the laser beams, is
\begin{equation}
2 h\nu =
\sqrt{ p_f^2c^2 + (m c^2 + 2h\nu_o)^2} -
\sqrt{ p_i^2c^2 + (m c^2)^2} .
\end{equation}
\noindent
where the rest mass of the atom in the initial
state is $m$ and $\nu_o$ is the unperturbed transition frequency.
Expanding, we obtain 
\begin{equation}
\nu  = 
\nu_o
+\underbrace{\frac{({\bf k}_1+{\bf k}_2)\cdot{\bf p}_i}{4\pi m}
\left(1-\epsilon\right)}
	_{\textstyle \Delta\nu_{D1}}
+\underbrace{\frac{\hbar({\bf k}_1 + {\bf k}_2)^2}{8\pi m}
\left(1-\epsilon\right)}
	_{\textstyle \Delta\nu_{R}}
-\underbrace{\frac{\nu_o p_i^2}{2(mc)^2}}_{\textstyle \Delta\nu_{D2}}
+O {\left( \frac{h \nu}{mc^2}\right)^3}.
\end{equation}
Here $\Delta\nu_{D1}$ and $\Delta\nu_{D2}$ are the first and second
order Doppler shifts, respectively, $\Delta\nu_R$ is the recoil shift,
and $\epsilon=2h\nu_o/mc^2=1.1\times 10^{-8}$ is a relativistic
correction which accounts for the mass change of the atom upon
absorbing energy $2h\nu_o$.  For hydrogen in the submillikelvin
regime, $\Delta\nu_{D2} \ll 1$~Hz, and can be neglected.  In the
Doppler-sensitive configuration, ${\bf k}_1={\bf k}_2$ and
$\Delta\nu_R = 6.7$~MHz.  (All frequencies are referenced to the
243~nm laser source).  At a temperature of $50~\mu$K, $\Delta
\nu_{D1}\sim 2.6$~MHz.  In the Doppler-free configuration, ${\bf
k}_1=-{\bf k}_2$, and there is no recoil or first order Doppler
broadening.  Doppler-free excitation is achieved by retro-reflecting
the laser beam.  Nevertheless, in this configuration the
Doppler-sensitive line is also excited, with the atom absorbing two
photons from a single laser beam.

\begin{figure}[t]
\centering\epsfig{file=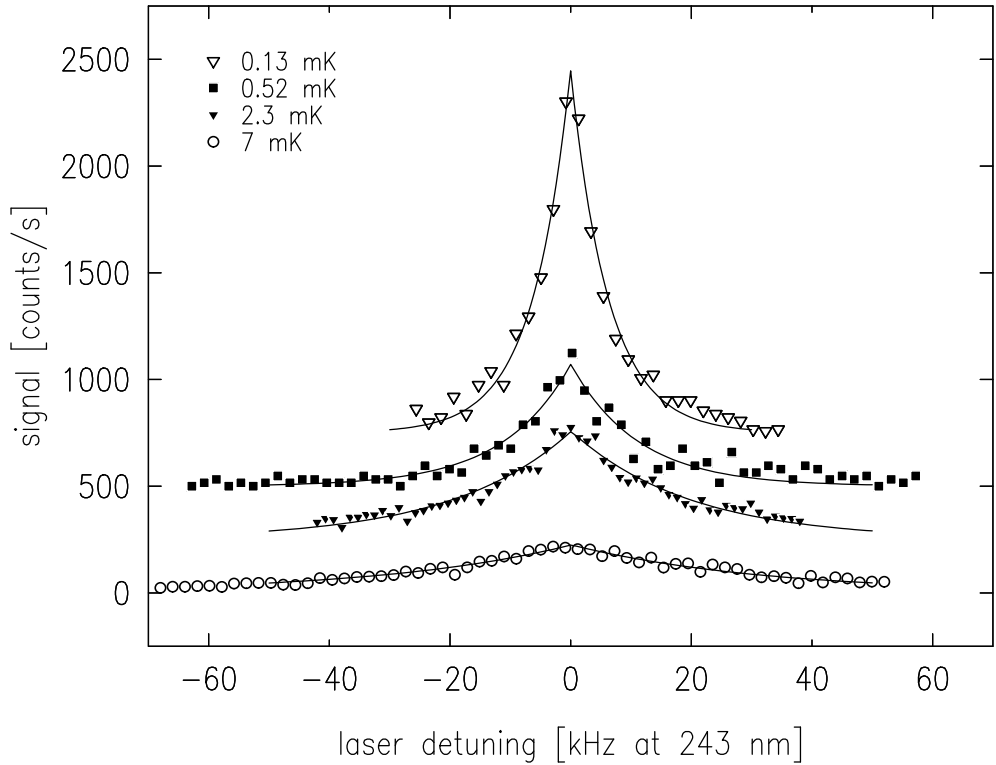}
\caption{Spectra of low density samples
at a series of temperatures.  The linewidth is given by the finite
interaction time as an atom traverses the Gaussian laser beam.  The
exponential lineshape arises from an average over a Maxwell-Boltzmann
distribution of velocities.  As the sample cools, the characteristic
velocity decreases and the line narrows $\sim \sqrt{T}$.  
The laser power was about 7~mW, and densities were in the 
range $10^{12} \sim 10^{13} {\rm cm}^{-3}.$}  
\label{tof.spec.fig}
\end{figure}

\begin{figure}[t]
\centering\epsfig{file=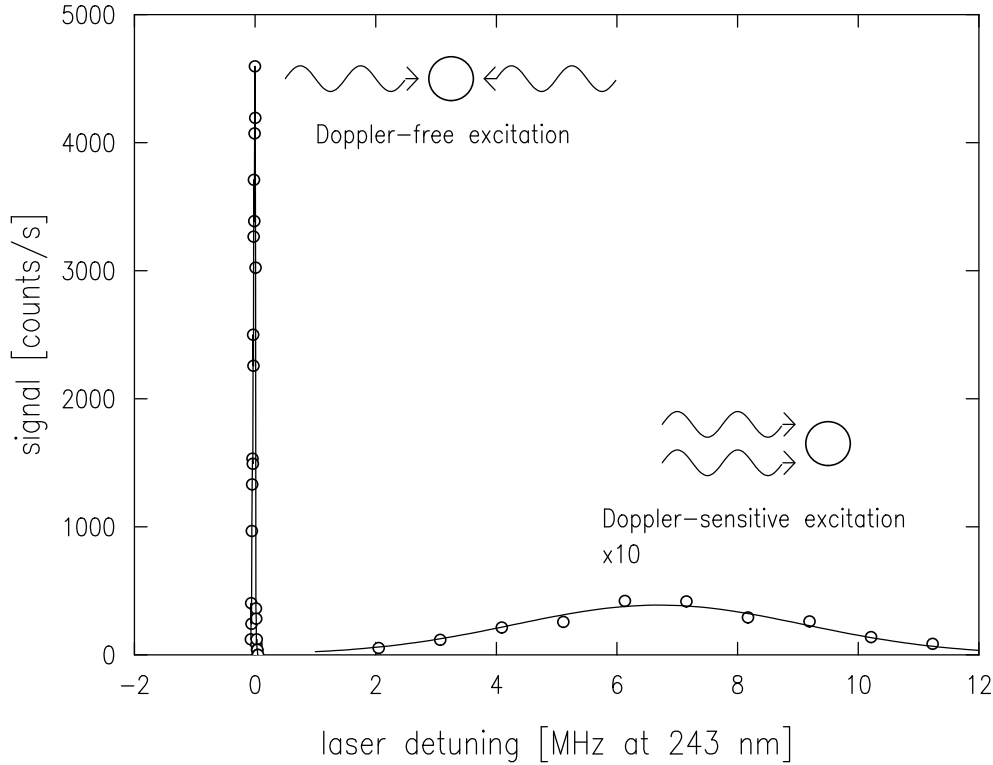}
\caption{ Composite $1S$-$2S$ two-photon spectrum of trapped hydrogen
before condensation.  The intense, narrow peak arises from absorption
of counter-propagating photons by the normal gas, and exhibits no
first-order Doppler broadening.  The wide, low feature on the right is
from absorption of co-propagating photons.  The solid line is the
recoil-shifted, Doppler-broadened, Gaussian lineshape of the normal
gas corresponding to $T=40~\mu$K.  Zero detuning is taken for
unperturbed atoms excited Doppler-free.  All frequencies are
referenced to the $243$~nm excitation radiation.}
\label{panoramic.spec.fig}
\end{figure}

Far from quantum degeneracy, the lineshape
for Doppler-sensitive excitation is the familiar Gaussian curve
characteristic of a Maxwell-Boltzmann distribution.  The shape for
Doppler-free excitation, however, is quite different---a cusp-shaped
double exponential \cite{bbc79}: $I(\nu-\nu_o) \sim \exp(-\mid
\nu-\nu_o\mid /\delta_o)$.  The linewidth parameter $\delta_o$ is
determined by the time of flight of an atom across the laser beam:
$\delta_o = u/2 \pi d_o$ where $u = \sqrt{2 k_{\rm B} T/m}$ is the
most probable velocity and $d_o$ is the waist diameter of the laser
beam.  This expression is valid for an untrapped gas far from quantum
degeneracy.  It neglects collisions and other broadening mechanisms,
and the natural linewidth of 1.3~Hz.  An example of this Doppler-free
lineshape is shown in fig.\ \ref{tof.spec.fig}.  
A panoramic spectrum
showing the Doppler-free and the recoil-shifted Doppler-sensitive
lines is shown in fig.\ \ref{panoramic.spec.fig}.

If the atoms are confined in a radially harmonic trap, the
Doppler-free spectrum consists of a central line at frequency $\nu_o$
plus a series of sidebands spaced by twice the trap frequency, lying
under the exponential curve \cite{clz97}.  The intensity of the
sidebands is governed by the ratio of the atom cloud diameter to laser
beam diameter.  If this ratio is less than 1, only the cental line is
excited, an example of Dicke narrowing.

\section{EVAPORATIVE COOLING}
The physical principles and experimental considerations of
evaporative cooling have been described in detail by Ketterle and van
Druten \cite{kdr96}.  We summarize here some of the principal points.

Evaporative cooling occurs when highly energetic atoms are permitted
to escape from a trap at a rate that is kept sufficiently low for the
remaining gas to maintain a quasi-thermal equilibrium.  In this
situation the energy distribution is thermal except that it is
truncated at the depth of the trap, $V_{\rm trap}$.  The crucial
parameters governing evaporative cooling are:

{\it The elastic collision rate.}  This determines the rate of
thermalization.  The low temperature collision cross section in a Bose
gas is determined by the $s$-wave scattering length, $a$.  The elastic
collision cross section for identical particles is $\sigma _{\rm el} = 8
\pi a^2, $ and the elastic collision rate is $\Gamma_{\rm el} = \sqrt{2}
n\sigma_{\rm el}\bar{v}$, where $n$ is the density and $\bar{v} =
\sqrt{8k_{\rm B} T/\pi m}$.  Table~\ref{atomic.props.table} shows
values of $a$ and $\sigma_{\rm el}$ for hydrogen and a number of alkali
metal atoms.
\begin{table}[tb]
\centering\begin{tabular}{|c|c|c|c|c|c|c|}
\hline
species & $m$ (amu)  & $a$ (nm) & $\sigma_{\rm el}$ (\rm cm$^2$) & 
$T_c$ ($\mu$K) & $n_c$ (cm$^{-3}$) & $N_c$ \\
\hline
H \cite{fkw98} & 1 & 0.0648 \cite{jdk95} & $1.06\times 10^{-15}$ & 
50 & $1.8\times 10^{14}$ & $10^9$ \\
Li \cite{bsh97} & 7 & -1.45 \cite{ams95} & $5.2\times 10^{-13}$ & 
0.30 & $1.5 \times 10^{12}$ & $10^3$ \\
Na \cite{sac98} & 23 & 2.8 \cite{twj96} & $1.90\times 10^{-12}$ & 
2.0 & $1.5\times 10^{14}$ & $10^7$ \\
Rb \cite{bgm97} & 87 & 5.4 \cite{vtf97} & $7.3\times 10^{-12}$ & 
0.67 & $2.2 \times 10^{14}$ & $2\times 10^{6}$  \\
\hline
\end{tabular}
\caption[comparison of BEC experiments] {Comparison of the atomic
species that have been Bose condensed.  For each cited experiment,
$a$ is the scattering length,
$\sigma_{\rm el}$ is the elastic scattering cross section, $T_c$ is the
temperature of the onset of BEC,  $n_c$ is the
density of the thermal cloud at the onset, and $N_c$ is the
maximum number of condensate atoms.}

\label{atomic.props.table}
\end{table}
Hydrogen's anomalously small scattering length is conspicuous.  In
consequence, its elastic scattering cross section is smaller than that
of the alkali metal atoms by a factor of a thousand or more,
and evaporative cooling proceeds at a correspondingly low rate.

{\it The cooling path.}  The ratio of trap depth to thermal energy,
$\eta \equiv V_{\rm trap}/k_{\rm B} T$ is an important parameter in
determining the most efficient evaporation path.  The rate at which
atoms with enough energy to escape from the trap are generated is
approximately proportional to $\eta \exp(-\eta)$.  Each atom carries
away energy $\sim \eta k_{\rm B} T$.  If $\eta$ is large, each
escaping atoms carries away a great deal of energy, enhancing the
efficiency of cooling,  but because the number of these atoms is
small, their evaporation proceeds slowly.  If $\eta$ is too small, the
atoms can escape too rapidly for the system to thermalize, and the
efficiency falls.

{\it Trap geometry.}  In a square-well potential, the density of a
trapped gas is independent of its energy, and the elastic collision
rate $\Gamma_{\rm el} = \sqrt{2} n \bar{v} \sigma$ decreases as $ \bar{v}
\sim \sqrt{T}$: evaporation slows as the temperature falls.  In a
trap, however, as the cloud cools the atoms are compressed into the
region of lowest potential, and the density of the gas increases as
its temperature falls.  As a result, in a harmonic trap the
evaporation rate increases with falling temperature as $\Gamma_{\rm el}
\sim T^{-1}$.  In a quadrupole trap, $\Gamma_{\rm el}\sim T^{-5/2}$.
Consequently, the functional form of the trap shape is a critical
design factor.  Because the evaporation rate in a trap increases as
the temperature falls, one can achieve conditions of runaway
evaporation in which evaporation, once started, can proceed faster and
faster.

{\it Loss processes.}  If there were no loss mechanisms in the trapped
gas, the time for evaporation could be made as long as desired and the
magnitude of $\Gamma_{\rm el}$ would be unimportant.  However, loss
processes are inevitable.  In experiments with alkali metal atoms the
principal loss is scattering of atoms out of the trap by collisions
with the warm background gas in the cell.  The loss rate is
independent of the density of the trapped gas.  (At very high density,
however, loss due to 3-body recombination plays the limiting role.)
In the cryogenic environment of atomic hydrogen experiments, the major
loss process is dipolar decay at a rate proportional to density.
Because both the dipolar decay rate and $\Gamma_{\rm el}$ vary linearly
with density, the conditions for runaway evaporation are not achieved.

\section{EVAPORATION TECHNIQUES}

In the initial demonstration of evaporative cooling the atoms were
permitted to escape over a saddle point in the magnetic field at one
end of the long cylindrical trap.  Evaporation is forced by lowering
this axial confinement field while simultaneously holding the radial
confinement fields fixed.  Energetic atoms are able to escape out the
end of the trap.  With this method, called saddle point evaporation,
it was possible to achieve conditions close to BEC in hydrogen, but
the cooling power was not adequate to cross the barrier.

The inefficiency of the evaporation process has been explained by
Surkov \etal\ \cite{sws96}.  In saddle point evaporation atoms escape
only along the z-axis.  For an atom to escape, it must have a
sufficient energy in the axial degree of freedom, $E_z\geq V_{\rm
trap}$, where $V_{\rm trap}$ is the trap depth as set by the
saddle point potential.  Because only the $z$-motion is involved, the
evaporation is inherently one-dimensional.  If the axial and radial
motion mix rapidly, then all atoms with total energy $E\geq
V_{\rm trap}$ can promptly escape.  At high energy this mixing takes
place in our trap because of the coupling of radial and transverse
co-ordinates.

As the energy decreases and the trap becomes more harmonic, the mixing
time lengthens.  When it becomes comparable to the collision time, the
evaporation rate falls.  This is because the collision of an energetic
atom generally transfers it to lower energy: the atom is knocked back
into a trapped energy regime.  Mixing is governed by the adiabaticity
parameter, $\Phi=\dot{\omega}_\rho/\omega_\rho^2$, which quantifies
the fractional change in the radial oscillation frequency per
oscillation period as the atom moves along the trap axis (see eq.\
\ref{osc.freq.eqn}).  For $\Phi \ll 1$, the probability of transfering
the energy from radial to longitudinal during one radial oscillation
is $\sim \Phi$.  Using $\dot{\omega}_\rho=(\drm \omega_\rho/\drm
z)(\drm z/\drm t)$ we obtain
\begin{equation} 
\Phi=v_z\: \frac{\beta z \sqrt{m}}{\alpha \sqrt{\beta z^2 + \theta}}.
\end{equation} 
For our trap at the threshold for BEC, $\Phi \sim 10^{-3}.$ However,
the probability for scattering during a radial period,
$2\pi\Gamma_{\rm el}/ \omega_{\rho}$, is about ten times higher.
Hence, the evaporation is essentially one-dimensional.  Surkov \etal\
have shown that in this situation the cooling rate decreases by a
factor of about $4 \eta$ compared to the rate at which evaporation
would proceed in three dimensions \cite{sws96}.  For hydrogen, such a
decrease in the already low evaporation rate proves fatal.  The
effects of one-dimensional evaporation have been studied by Pinske
\etal\ \cite{pmw98}.

The bottleneck of saddle point evaporation is avoided by the technique
of radiative evaporation, ``rf evaporation,'' originally proposed by
Pritchard \cite{phm88}, which permits evaporation in three dimensions.
A radio-frequency oscillating magnetic field drives transitions
between the trapped state and some other (untrapped) hyperfine
sub-level, causing the atom to be ejected from the trap.  In hydrogen
the trapped hyperfine state is ($F=1,m=1$), and the rf transition is
predominantly to the state (1,0).  The resonance occurs for atoms in a
region of the trap where the field magnitude $B_0$ satisfies the
resonance condition $\hbar \omega_{rf}=\mu_{\rm B}B_0$.  Here
$\mu_{\rm B}B_0$ is the Zeeman splitting between the hyperfine
sub-levels.  The transition matrix element is $\hbar \Omega_R =
\mu_{\rm B} B_{\perp}/\sqrt{2}$, where $B_{\perp}$ is the amplitude of
the rf field perpendicular to $B_0$.  The probability of an atom
experiencing a hyperfine sub-level transition as it traverses the
resonance region can be estimated using the Landau-Zener theory.  The
trapping field is assumed to vary linearly with gradient $B^\prime$ in
the vicinity of the resonance, and the atom traverses the region at
speed $v$.  The probability of a two-level atom making a transition as
it traverses the resonance region is $p=1-\exp(-\zeta)$ where
$\zeta=2\pi\hbar\Omega_R^2/\mu_{\rm B} B^\prime v$ \cite{rub81}.  The
hydrogen $F=1$ state is actually a three level system.  Vitanov and
Suominen have solved the multilevel Landau-Zener problem \cite{vsu97},
and we apply their results to our situation.  The probability $P_m$
that an atom originally in the state (1,1) emerges in the state
$(1,m)$ is
\begin{eqnletter}
P_1 &=& (1-p)^2 \\
P_0 & = & 2(1-p)p \\
P_{-1} &=& p^2
\end{eqnletter}
For large $\zeta$ the atom absorbs two rf photons and emerges in the
(1,-1) state, which is ejected.  In our experiment $\zeta$ is small,
and there is only a small probability of leaving the trapped (1,1)
state.  Transitions are primarily to the (1,0) state.  Atoms in this
state simply fall out of the trap.

For small $\zeta$, the probability of being ejected during one radial
oscillation is $\sim 4\zeta$. As in the previous discussion on radial
and longitudinal mixing, scattering sets a lower limit of $\zeta \geq
10^{-3}$.  This requires a Rabi frequency of $\Omega_R\geq 2\pi\times
1$~kHz or an rf field strength $B_\perp\geq 10^{-7}$~T.  (In other
atomic systems dipolar relaxation can impose a more stringent
requirement on the transition rate \cite{kdr96}.)

To implement rf evaporation in a cryogenic environment methods had to
be developed to eliminate eddy current losses and rf shielding by the
cell.  This was accomplished using a plastic cell design in which
heat transport is provided by a superfluid helium jacket around
the cell.  The rf field amplitude is typically $7 \times 10^{-7}$~T,
and fields can be applied with frequencies up to 46~MHz.  RF
evaporation is switched on at a trap depth of about 1.1 mK, where the
sample temperature is typically 120~$\mu$K.

In addition to driving evaporation, the rf field is used to find the
temperature of the sample.  The temperature is measured by sweeping
the rf resonance through the trap and measuring the atom ejection rate
as a function of frequency \cite{phm88}.

The field at the bottom of the trap, the axial bias field
$\theta/\mu_{\rm B}$, is a critical parameter because it determines
the curvature of the potential minimum and provides the zero point for
determining the trap depth.  If the field is too large the trap is
harmonic, the effective volume is large, and the density is small. If
the bias field is too low, the condensate is strongly confined and
suffers a high dipolar decay rate.  The optimum bias field is
approximately $k_{\rm B} T/\mu_{\rm B}$.  In such a field the thermal
cloud is tightly confined, but the condensate can spread out due to
interactions.  In our experiments the bias field energy is about
$\theta \simeq (3/5)k_{\rm B} T$.  The bias field is measured by
applying the rf field at a low frequency and sweeping the frequency up
until atoms start to leave the trap.

We have neglected the effect of gravity.  For \Hup\ at a temperature
of $50~\mu$K the gavitational scale height $k_{\rm B}T/mg$ is 4~cm,
which is comparable to the vertical size of the cloud.  For higher
mass atoms, which condense at lower temperatures, gravity can be
important.  In such a case the surface of constant $B_0$ is no longer
an equipotential.  Evaporation occurs primarily from the bottom of the
cloud, and becomes one-dimensional \cite{kdr96}.

\section{COLD-COLLISION FREQUENCY SHIFT}

At low temperature, in the limit $a \ll \Lambda_{\rm T}$, only
$s$-wave collisions are important.  These collisions give rise to a
mean interaction energy, and they introduce frequency shifts into
radiative transitions.  Such effects can be analyzed by kinetic theory
starting with the Boltzmann transport equation \cite{vks87,tvs92}, or
described by mean field theory based on the pseudopotential
\cite{pat72}.  Both approaches give the same result for a homogenous
system: the mean field energy of an atom in a gas with density $n$ is
\begin{equation}
\tilde{U} n = \frac{ 4 \pi \hbar^2 a}{m} n \times g_2(0) ,
\label{MeanFieldEnergy:eqn}
\end{equation}
where the density normalized second order correlation function
$g_2(\bf{x})$ is \cite{hov54} 
\begin{equation}
g_2 ({\bf x}) = \frac{1}{n N} \sum_{i \neq j} \left\langle \Psi \mid
\delta ( {\bf r_i} - {\bf r_j} - {\bf x})\mid\Psi\right\rangle .
\end{equation}  
Here $\Psi$ is the wave function for the system.  (For a Bose gas far
from degeneracy, $g_2(0) = 2$.)  Because the scattering lengths for
\oneSoneS\ and \oneStwoS\ collisions are not identical, the energy to
excite an atom to the $2S$ state from a gas of $1S$ atoms is shifted
by an amount \cite{footnote1}
\begin{equation}
h\Delta \nu_{\rm col} 
=\frac{4 \pi \hbar^2 n }{ m}(a_{1S-2S} - a_{1S-1S})g_2(0).
\label{eq:Ecol}
\end{equation}
The frequency shift $\Delta \nu_{\rm col}$ is known as a cold
collision frequency shift.  For a non-degenerate gas the two-photon
sum frequency is shifted by $\Delta \nu_{\rm col} =n\chi$, where $\chi
= 4 \hbar (a_{1S-2S} - a_{1S-1S})/m$. Once $\chi$ is known, the
density can be determined directly by measuring the frequency shift.
In addition, a measurement of $\chi$ can be used to check the
theoretical calculations of the scattering lengths.  The $1S$-$1S$
scattering length is known accurately from theory: $a_{1S-1S} =
0.0648$~nm \cite{jdk95}, and $a_{1S-2S}$ has also been computed:
$a_{1S-2S} = -2.3$~nm \cite{jdd96}.

The cold-collision shift of the \oneStwoS\ transition plays a key role
in our experiments, allowing us to measure the density rapidly {\it in
situ}.  Furthermore, because the density in a Bose-Einstein condensate
in \Hup\ is much higher than the density of the normal gas, the
cold-collision frequency shift provides an unmistakable signature of
the condensate.

\begin{figure}[t]
\centering\epsfig{file=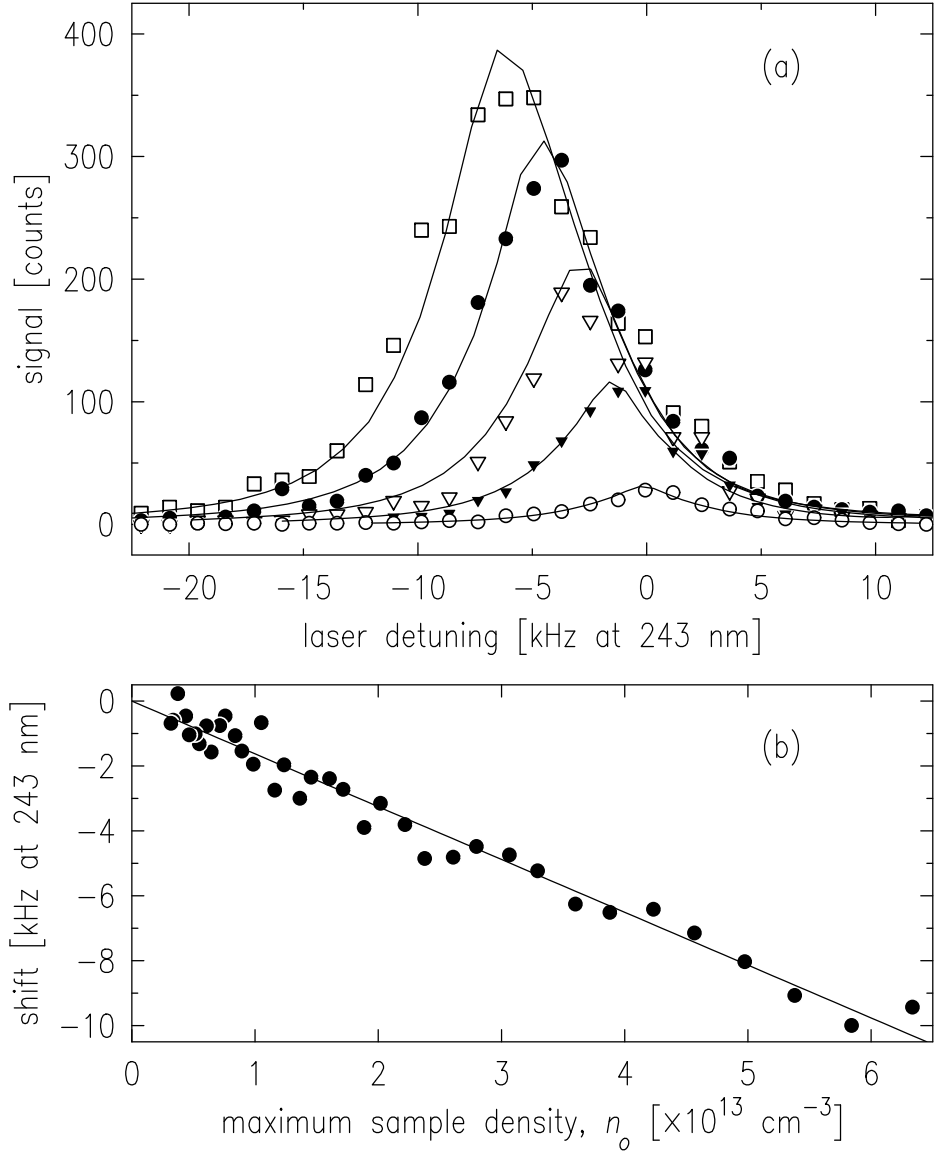}
\caption{ (a) Series of spectra of a single sample used for a
measurement of $\chi$.  The first scan is at the maximum density and
exhibits the largest red shift.  Subsequent scans, at lower densities,
are smaller and less shifted. Five of the forty spectra are shown. (b)
Shift of the spectrum as a function of sample density.  The density is
inferred from the integrated signal.  For this particular sample
$\chi= -3.30 \pm 0.6 \times 10^{-10} {\rm~Hz~cm}^3$.  From
\cite{kfw98}. }
\label{density.scans.fig}
\end{figure}

To measure the frequency shift parameter $\chi$, a series of line
scans were taken at different densities as shown in fig.\
\ref{density.scans.fig}a.  
The initial density was established by
monitoring the two-body dipolar decay rate, as described above.
During successive laser scans of a single trapped sample the density
decayed because of collisions with helium gas generated by heating due
to the laser.  The area under each photoexcitation curve is
proportional to the total number of atoms, making it possible to infer
the density for each scan.  The line center for each curve was
corrected for the effects of density inhomogeneities due to the
trapping potential.  From a plot of frequency {\em vs.}\ density, like
the one shown in fig.\ \ref{density.scans.fig}b, the value of $\chi$
can be determined.  From a series of such measurements taken at
different densities and temperatures, we obtained $\chi = -3.8 \pm 0.8
\times 10^{10}~n~{\rm Hz \, cm}^3$.  The theory of cold-collision
frequency shifts in an inhomogeneous system is not yet fully
understood, but assuming eq.\ \ref{eq:Ecol} is still valid, we deduce
$a_{1S-2S} = -1.4\pm 0.3$~nm, in fair agreement with the prediction
\cite{kfw98}.

\section{OBSERVATION OF THE CONDENSATE}

\begin{figure}[t]
\centering\epsfig{file=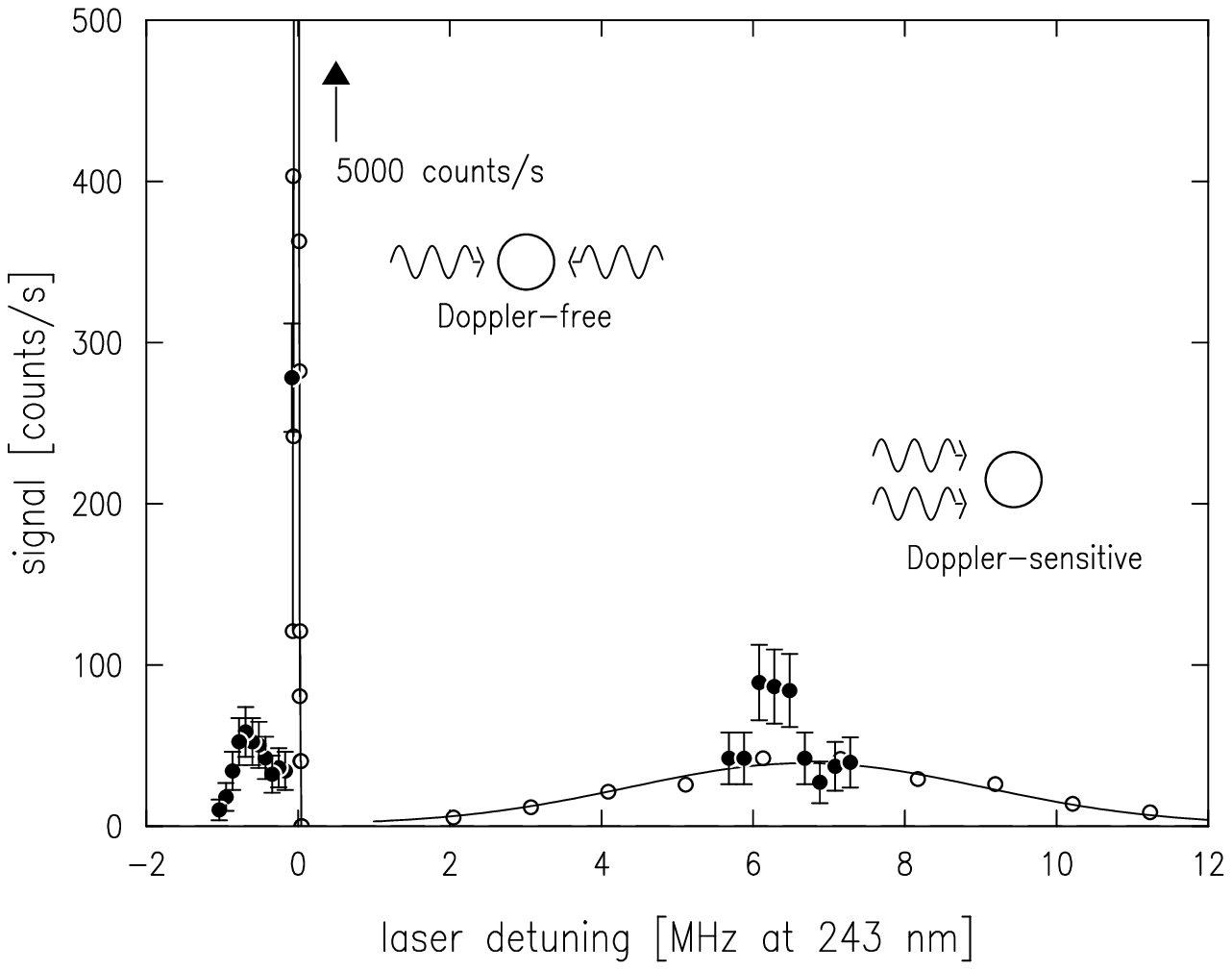}
\caption{Composite $1S$-$2S$ two-photon spectrum of trapped hydrogen
after condensation.  $\circ$--spectrum of sample without a condensate;
$\bullet$--spectrum emphasizing features due to a condensate.  The
high density in the condensate shifts a portion of the Doppler-free
line to the red.  The condensate's narrow momentum distribution gives
rise to a similar feature near the center of the Doppler-sensitive
line.  From \cite{fkw98}.}
\label{overall.spec.fig}
\end{figure}

Bose-Einstein condensation in hydrogen was achieved at a temperature
of about $50~\mu$K with a density of $1.8 \times 10^{14}~{\rm
cm}^{-3}$ \cite{fkw98}.  Its onset was revealed by unmistakable new
features in the spectrum, shown in fig.\ \ref{overall.spec.fig}.  
In
the Doppler-sensitive line there is a narrow peak, shifted somewhat to
the red of the line center, and there is a similar line to the red of
the Doppler-free line.  (Note: the Doppler sensitive peak was not
observed until after these lectures.)  The shift of these lines
reveals a density significantly higher than in the normal gas, as
expected for the condensate.

\begin{figure}[t]
\centering\epsfig{file=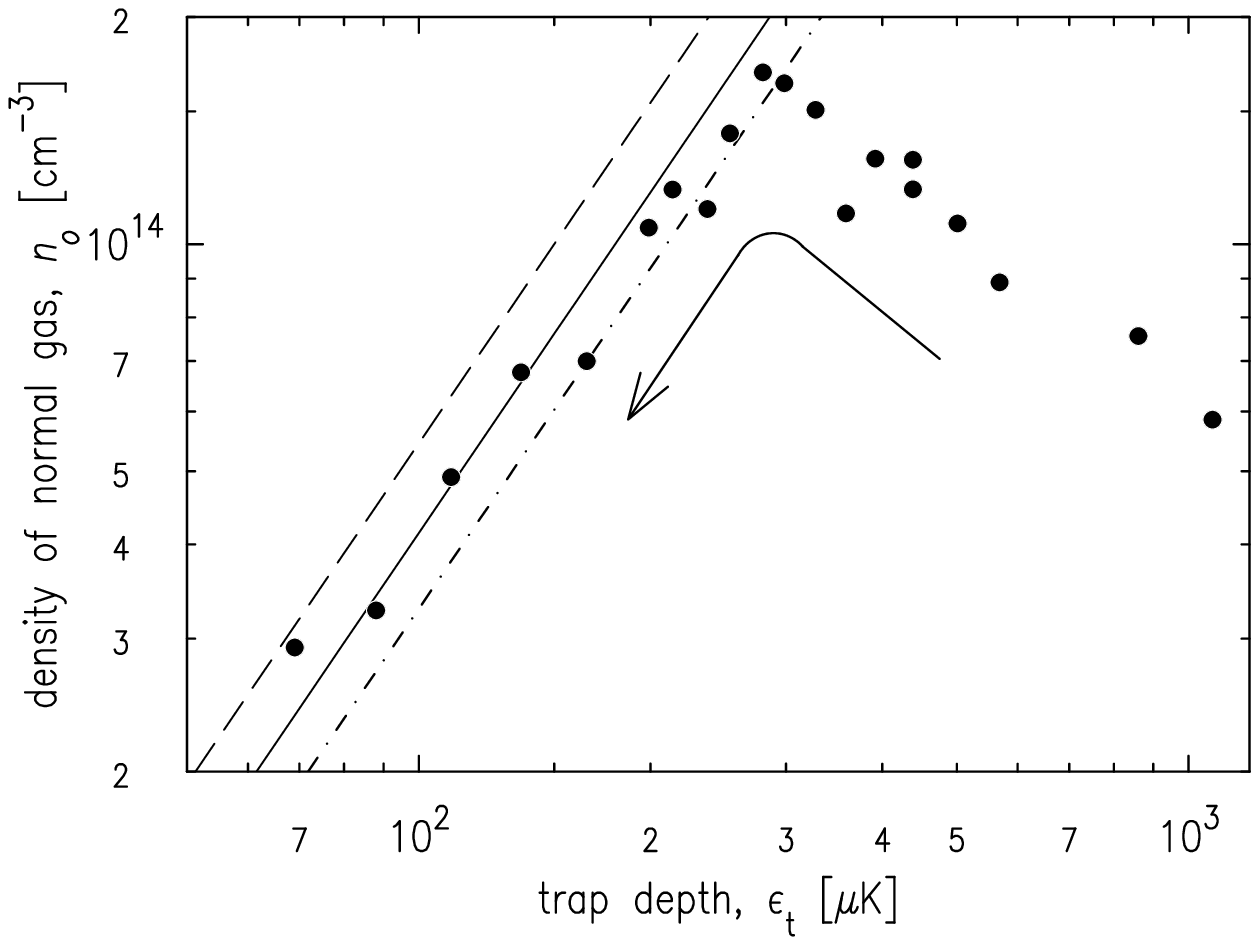}
\caption{Density of non-condensed fraction of the gas as the trap
depth is reduced along the cooling path.  The density is measured by
the optical resonance shift, and the trap depth is set by the rf
frequency.  The lines (dash, solid, dot-dash) indicate the BEC phase
transition line, assuming a sample temperature of (1/5th, 1/6th,
1/7th) the trap depth.  The scatter of the data reflects the
reproducibility of the laser probe technique and is dominated by
alignment of the laser beam to the sample.  From
\cite{fkw98}.}
\label{phasespace.fig}
\end{figure}

Before BEC was observed in the spectrum there were strong indications
that condensation was taking place from a study of the evolution of
the density of the thermal cloud with decreasing temperature.  The
density of the non-condensed gas fraction was determined from the
cold-collision frequency shift; the temperature was inferred from the
trap depth, set by the frequency of the rf signal.  A plot is shown in
fig.\ \ref{phasespace.fig}.  
In these observations the value of $\eta$
typically varied between 7 and 5 with decreasing temperature.  The
solid line in fig. \ref{phasespace.fig} is the BEC boundary for $\eta
= 6$.  Because the normal gas cannot exist to the left of the
boundary, the density simply falls along the boundary.  Once the
system reaches the boundary, as the temperature is further reduced the
normal atoms are forced into the condensate.  Because the observations
are at a laser frequency tuned so that only atoms from the normal gas
are excited, the condensed atoms, which are at such a high density
that they are frequency-shifted out of range, are not observed.
Consequently, the density of the thermal cloud merely tracks the BEC
line.

\begin{figure}[t]
\centering\epsfig{file=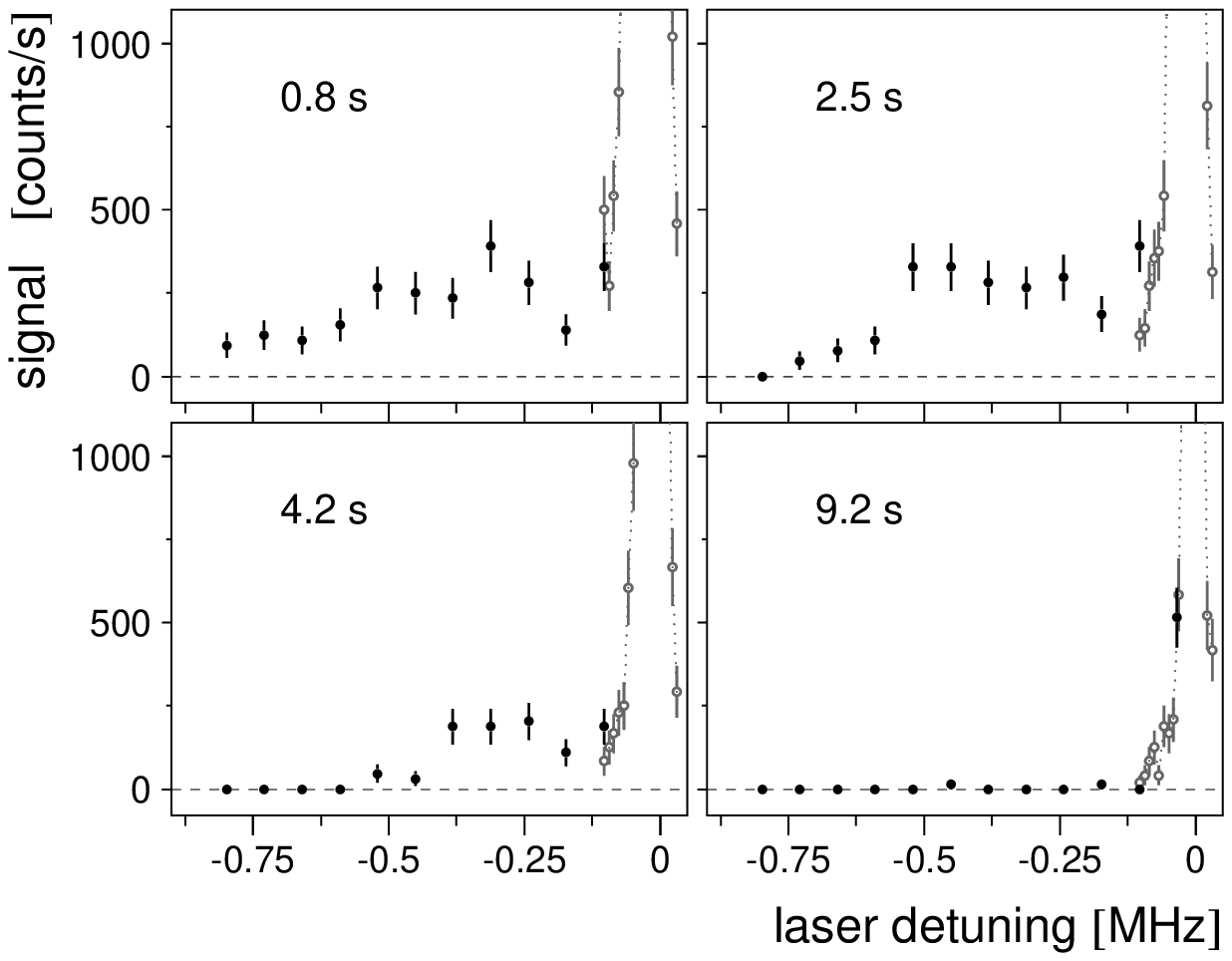}
\caption{  Time evolution of the condensate.  Each
spectrum is obtained in 0.67~s.  As the condensate disappears the
spectrum narrows, indicating a reduction in density, and gets weaker,
indicating a reduction in condensate population.  The last panel shows
the background count rate.  From
\cite{fkw98}.}
\label{condensate.evolution.fig}
\end{figure}

Because of the condensate's high density, its dipolar decay rate is so
high that, in isolation, it would disappear in about one second.
However, because the condensate is continuously fed by the normal gas,
its lifetime is several seconds.  Its time evolution is shown in the
series of spectral scans shown in fig.\
\ref{condensate.evolution.fig}.  
As the density in the condensate
decreases, the red shift decreases and the total signal becomes
smaller, finally vanishing.

\section{PROPERTIES OF THE CONDENSATE}

\begin{figure}[t]
\centering\epsfig{file=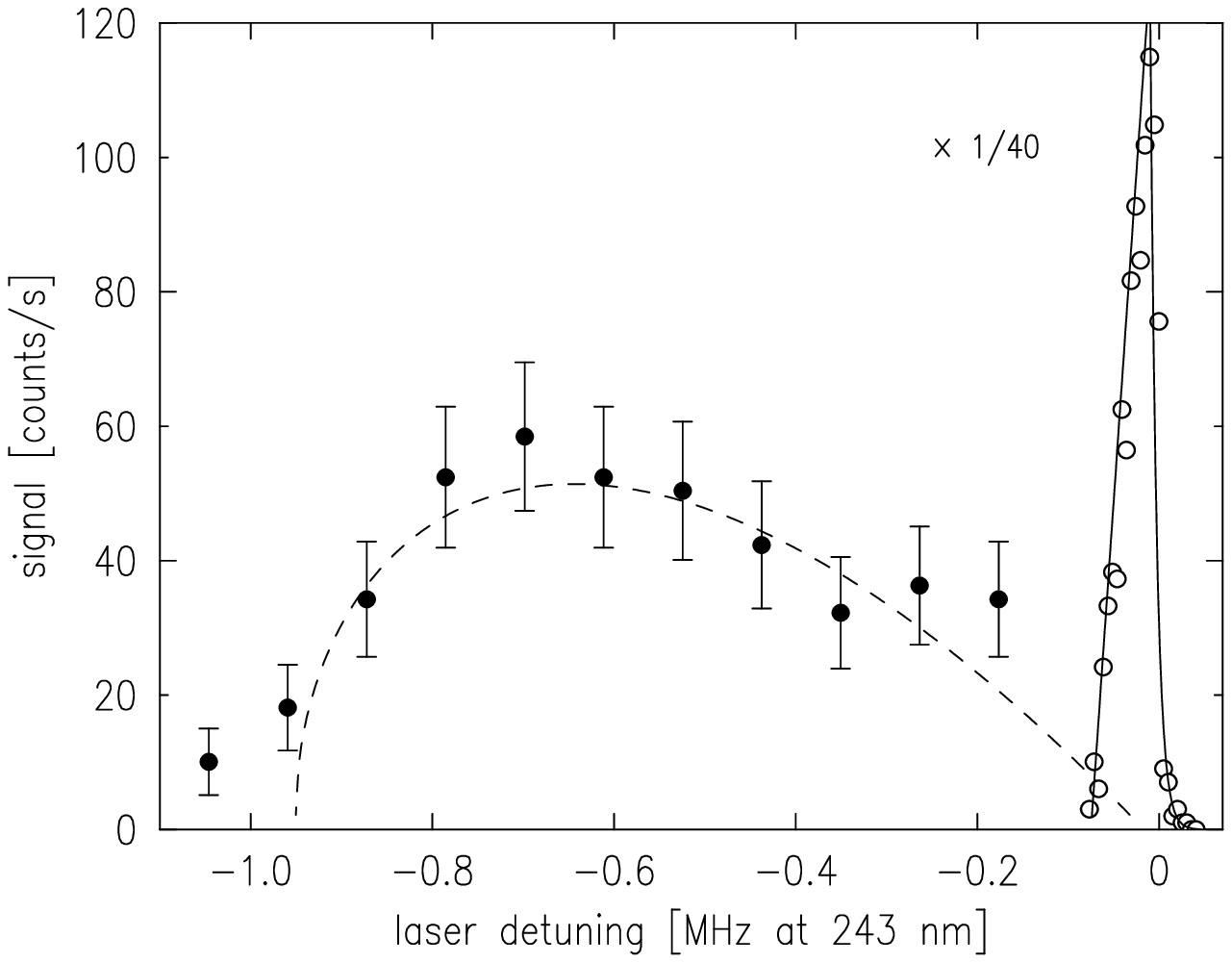}
\caption{Doppler-free spectrum of the condensate (broad feature) and
normal gas (narrow feature).  The dashed line is proportional to the
number of condensate atoms at a density proportional to the detuning,
for an equilibrium density distribution with peak density
$n_{o,c}=4.8\times 10^{15}~{\rm cm}^{-3}$ in a parabolic trap.  From
\cite{fkw98}.}
\label{cond.spec.fig}
\end{figure}

The peak condensate density $n_{o,c}$ is found from the red cut-off of
the spectrum.  As shown in fig.\ \ref{cond.spec.fig},
$n_{o,c}=4.8\pm1.1\times 10^{15}~{\rm cm}^{-3}$.  The peak mean field
energy of an atom in the condensate, $\tilde{U} n_{o,c}/k_{\rm B} = 4
\pi\hbar^2 a n_{o,c}/k_B m = 1.9~\mu$K, is much larger than the energy
interval between the radial vibrational states of the trap,
$\hbar\omega{\rho}/k_{\rm B} = 190$~nK.  Consequently, the shape of
the condensate is determined by the mean field energy rather than the
wave function of the trap's ground state.  The density profile, in the
Thomas-Fermi approximation, is $n({\bf r}) = n_{o,c} - V({\bf
r})/\tilde{U}$.

With the Thomas-Fermi wavefunction, the total number of atoms in the
condensate is found by integrating the density over the volume of
the condensate.  The result is
\begin{equation}
N_c = \frac{2^{9/2} \pi\; \tilde{U}^{3/2}\; n_{o,c}^{5/2}}
{15\omega_\rho^2\; \omega_z \; m^{3/2}} = (1.1\pm 0.6) \times
10^9~{\rm atoms}.
\label{Nc:eqn}
\end{equation}
The oscillation frequencies (eqs.\ \ref{osc.freq.eqn}) are
$\omega_\rho=2\pi\times 3.90\pm0.11$~kHz, and $\omega_z=2\pi\times
10.2$~Hz.  The diameter of the condensate is $15~\mu$m and its length
is $5$~mm.  The huge aspect ratio, $\sim 400$, gives the condensate a
thread-like shape.

The fraction of atoms in the condensate is small because the
condensate is rapidly depleted by dipolar relaxation.  The condensate
size is limited by the low evaporative cooling rate in the normal gas,
which supplies cold atoms to the condensate \cite{hks93}.  The
fraction, $f = {N_c}/( N_n + N_c)$, can be found from the integrated
area of the normal and condensate Doppler-free spectra, taking into
account that although the entire condensate is in the laser beam, only a
portion of the normal gas interacts with the laser.  The condensate
fraction can also be found by comparing the number of condensate
atoms, determined from $n_{o,c}$ and the trap geometry, with the
number of normal atoms found by integrating the Bose occupation
function weighted by the trap density of states.  (The temperature is
measured from the Doppler-sensitive spectrum of the normal gas.)  The
trap shape cancels in the comparison.  The methods are in good
agreement at $f = 6_{-3}^{+6}\%$.  Both methods assume thermal
equilibrium, which may not be justified.

In calculating the density and size of the condensate, we have taken
$g_2(0)=1$ in computing the mean field energy (eq.\
\ref{MeanFieldEnergy:eqn}), but $g_2(0)=2$ in calculating $n_{c,o}$
(eq.\ \ref{eq:Ecol}).  Although this appears contradictory, it yields
a condensate fraction that is consistent with the observed intensity
ratio of the photoexcitation spectra and with the fraction predicted in
ref.\ \cite{hks93}.  If we consistently take $g_2(0)=1$, then eq.\
\ref{Nc:eqn} yields $N_c = 6\times 10^9$ and $f = 25\%$. If we
consistently take $g_2(0)=2$, then $N_c=3 \times 10^9$ and $f = 14\%$.
In a homogeneous condensate in thermal equilibrium one expects
$g_2(0)=1$, however this may not apply under our experimental
conditions.  The problem clearly requires further study.

\begin{figure}[t]
\centering\epsfig{file=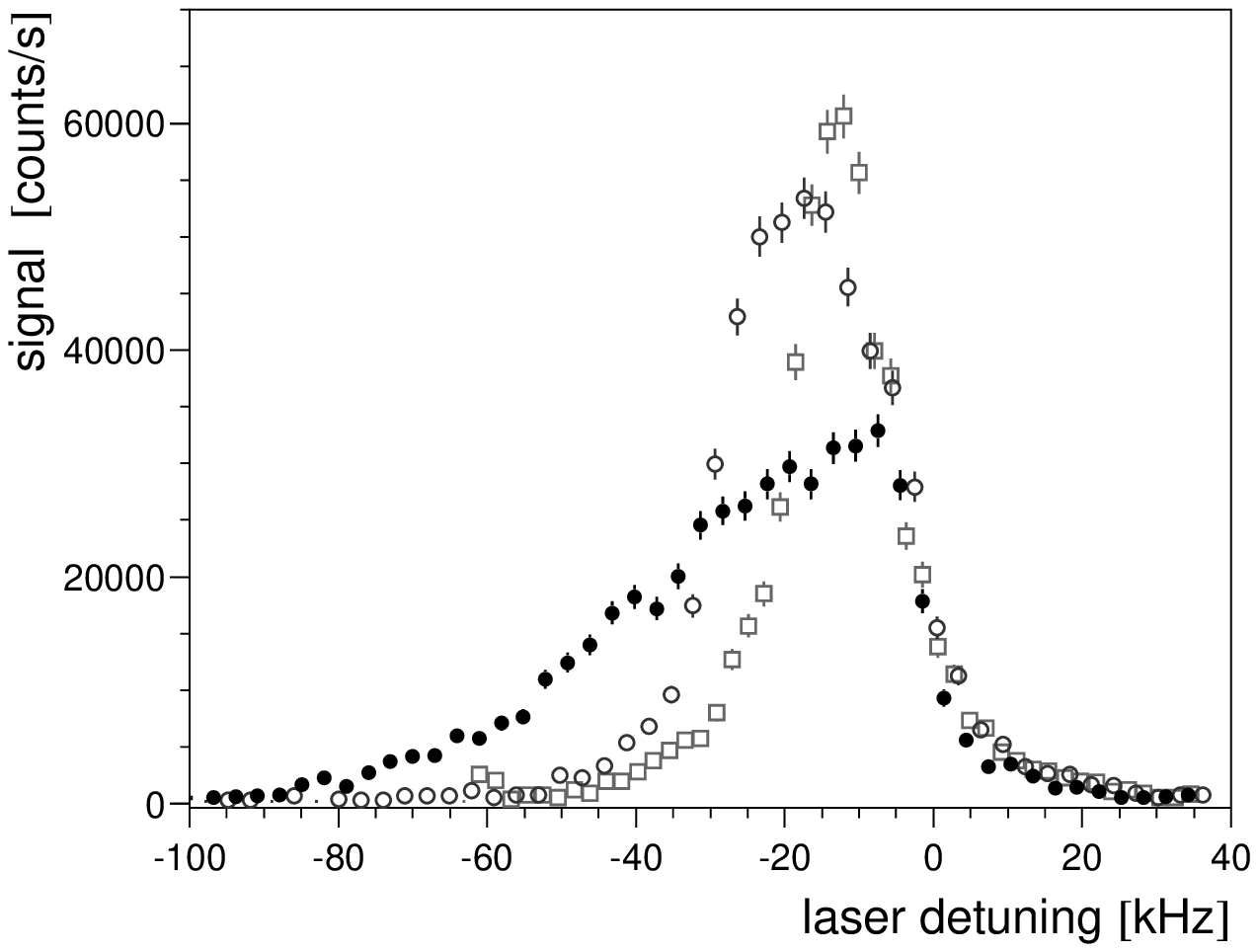}
\caption{Doppler-free spectrum of normal fraction above and below the
onset of BEC\@.  The symmetric spectrum (above $T_c$, open symbols)
suddenly becomes asymmetric (filled symbols) when the condensate
forms.  Temperatures for the three spectra are about $120~\mu$K (open
squares), $53~\mu$K (open circles), $44~\mu$K (filled circles).  From
\cite{fkw98}.}
\label{asymspec.fig}
\end{figure}

The Doppler-free lineshape of the normal gas displays puzzling
behavior at the condensation transition.  As seen in fig.\
\ref{asymspec.fig}, 
above the the transition temperature the line
displays a roughly symmetric shape, as expected for the density
distribution in the trap.  When the condensate is present, the line
develops a large asymmetry toward the red.  The ratio of normal to
condensate volumes is about $10^3$, so that this feature cannot be
explained simply by penetration of the normal gas into the condensate.

\section{PROSPECTS}

As one expects whenever a novel system is created, the observation of
BEC in hydrogen has led to some new questions.  In particular, the
nature of the photoexcitation spectrum from the condensate appears to
be more subtle than previously appreciated.  The cold-collision shift
has proven to be an invaluable diagnostic tool, but beyond that it has
led to an experimental value for an excited-state $s$-wave scattering
length---the first such determination to our knowledge.  One can
conceive of methods for extending such measurements to other excited
states.

Perhaps the most dramatic aspect of the hydrogen condensate is its
size,  more than thirty times larger than previous condensates,
with prospects for large improvements. Thus, hydrogen should be a
natural candidate for any application that requires an intense source
of coherent atoms.

The techniques for achieving BEC in hydrogen evolved over a long time,
and the apparatus reflects a great deal of history.  Vast improvements
would be possible if one were to start from scratch.  In particular,
the detection efficiency for \Lalpha\ photons is only about $2 \times
10^{-5}$, the chief loss in signal being due to the low optical
collection solid angle, $1.6 \times 10^{-2}$~sr.  An apparatus
designed for optical access would provide a much higher signal rate,
permitting a more precise study of the condensate's properties and its
dynamical behavior.

The size of the condensate is currently limited by the initial density
of hydrogen that can be loaded into the trap.  Improvements should be
possible.  On a more speculative note, if it were possible to increase
the thermalization rate of a hydrogen sample by introducing an
impurity atom into the trapped hydrogen gas, a major increase in the
condensate size might be achieved.

\section{Acknowledgements}
Many people played important roles in the MIT work on spin-polarized
hydrogen over the years.  We especially wish to acknowledge those who
contributed directly to the trapping and spectroscopy experiments:
Claudio L. Cesar, John M. Doyle, Harald F. Hess, Greg P. Kochanski,
Naoto Masuhara, Adam D. Polcyn, Jon C. Sandberg, and Albert I. Yu.

This research is supported by the National Science Foundation and the
Office of Naval Research.  The Air Force Office of Scientific Research
contributed in the early phases.  L.W. acknowledges support by
Deutsche Forschungsgemeinschaft.  D.L. and S.C.M. are grateful for
support from the National Defense Science and Engineering Graduate
Fellowship Program.


\begin{thebibliography}{000}
\bibitem{gterms.ref} In evaluating $g$ from \cite{lsv86} and
\cite{skv88}, we have taken $g=2(G_{dd\rightarrow aa}+G_{dd\rightarrow
ad}+G_{dd\rightarrow ac})+ G_{dd\rightarrow cd}+ 2G_{dd\rightarrow cc}$
using the low field and zero temperature values for the rate
constants.  When one atom changes hyperfine state ($F=1 \rightarrow
F=0$) both atoms are lost from the trap because the hyperfine energy
(71~mK) is much greater than the trap depth.
\bibitem{footnote1} A different expression was presented in the
lecture, and this expression requires some justification.  It is not
obvious that the \oneStwoS\ scattering leads to the same expression for the
mean field energy as \oneSoneS\ scattering.  The reason has to do with the
nature of excitation, in which not one particular $1S$ atom is excited
but every $1S$ atom is given a small component of $2S$ excitation.
Details will be published elsewhere.
\end{thebibliography}
\end{document}